\documentclass{iaus}
\usepackage{graphicx}

\title[UTM] 
{UTM, a universal simulator for lightcurves of  transiting systems}

\author[Hans J. Deeg]   
{Hans Deeg
}

\affiliation{Instituto de Astrof\'\i sica de Canarias, C. Via Lactea S/N, 38205 La Laguna, Tenerife, Spain \\ email: {\tt hdeeg@iac.es}}

\pubyear{2008}
\volume{253}  
\pagerange{1 -- 4}
\jname{Transiting Planets }
\editors{A.C. Editor, B.D. Editor \& C.E. Editor, eds.}
\begin{document}

\maketitle

\begin{abstract}
The Universal Transit Modeller (UTM) is a light-curve simulator for all
kinds of transiting or eclipsing configurations between arbitrary
numbers of several types of objects, which may be stars, planets,
planetary moons, and planetary rings. Applications of UTM to date have
been mainly in the generation of light-curves for the testing of
detection algorithms. For the preparation of such test for the Corot Mission, a special version has been used to
generate multicolour light-curves in Corot's passbands.  A separate
fitting program, UFIT (Universal Fitter) is part of the UTM
distribution and may be used to derive best fits to light-curves for
any set of continuously variable parameters. UTM/UFIT is written in
IDL code and its source is released in the public domain under the GNU General Public
License.

\end{abstract}

\firstsection 
\section{Overview}
A number of investigations related with transiting extra-solar planets
depend on a reliable modelling of the light-curves that occulting system may exhibit. In this paper, we present the UTM package, which
is short for Universal Transit Modeller. Its major strength is its
flexibility which allows the modelling of synthetic light-curves of a
very varied range of transiting configurations. These may range from
simple planet-star transits to situations with planets orbiting close
double stars; but UTM may also model the
effects of planetary moons or rings onto a transit signature. Applications of UTM to date have been the fitting of
model-curves to observed data (\cite[Deeg \& Garrido 2001)]{deeg+01} and the generation of light-curves for the evaluation of detection
algorithms (\cite[{Aigrain} \& {Favata} 2002]{aigrain+02}, \cite[{Aigrain} \& {Irwin} 2004]{aigrain+04}, \cite[{Carpano}, {Aigrain} \& {Favata} 2003]{carpano+03}, \cite[{Carpano} \& {Fridlund} 2008]{carpano08}). For the Corot mission, UTM has been used in simulations of transits of Earth-like planets in the face of stellar microvariability (\cite[{Bonomo} \& {Lanza} 2008]{bonomo08}) and for the preparation of several 'blind tests' to determine the mission's capability of transit detection, with detections from monocolour light-curves discussed in \cite[{Moutou} {et~al.}(2005)]{moutou+05} and  multicolour light-curves in \cite[{Moutou} {et~al.}(2007)]{moutou+07}, with a further test about the detectability of 'Super-Earths' currently being under way.

\begin{figure}[t]
\begin{center}
 \includegraphics[width=11cm]{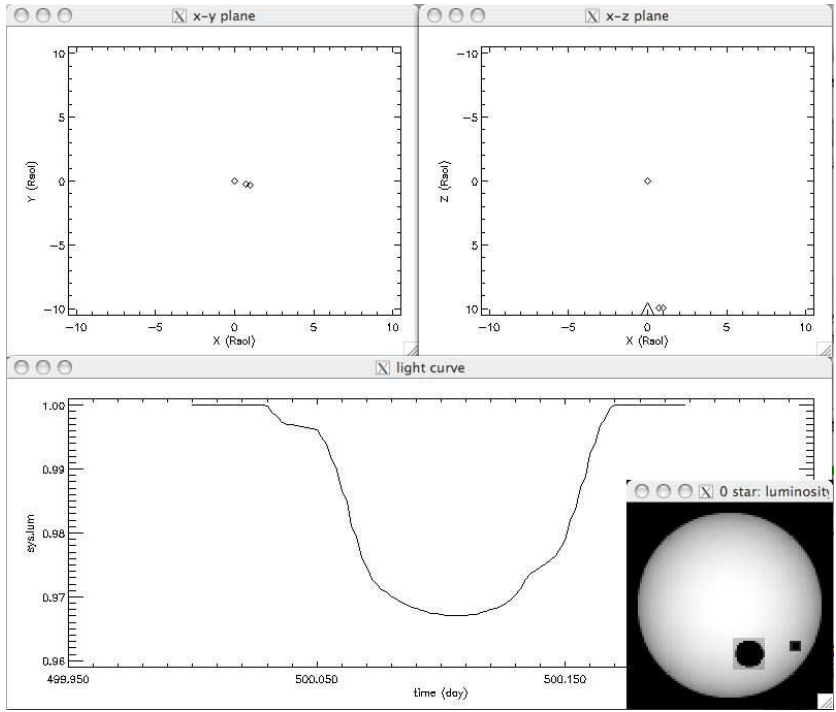} 
 \caption{Visualisations shown by UTM for the case of a transit of a planet with a moon across a star:  in the upper center and right are displays of objectsÕ positions in the x-y 'sky-plane' (as seen by observer) and in the x-z plane (as seen from ÔtopÕ, with the triangle at x=0 indicating the direction to the observer). The three diamonds correspond to the three objects. The lower panel shows the output light-curve at the end of the simulation, were the offset between the lunar and the planetary transit is visible. The insert in the lower right shows the surface of the star in a moment during the planet-moon transit. While running UTM, these windows show a continuously updated animation.}
   \label{fig_utmview}
\end{center}
\end{figure}

UTM is a light-curve modeller for all kinds of transiting or eclipsing
configurations between an arbitrary number of different object-types,
which may be stars, planets, planetary moons, and planetary rings. All input to UTM is controlled through setup files with a simple syntax. UTM calculates the orbital positions of the objects at each time-point and derives the
resulting brightness. The simulations may be obtained for equidistant time-points or for time-points taken from an
input file (e.g. from an existing light-curve). There are several levels of visualisations available, with animations of the bright objects of the system, of the positions of objects, and of the resultant light-curve (See Fig.~\ref{fig_utmview}), as well as a fast 'quiet mode'. Model light-curves may be produced in units of the
model-system's luminosity, as magnitude variations, or as relative
flux variations, and are either saved directly; after the adding of random noise; or after having been added to an existing light-curve.  UTM generates monocolor light-curves only, but a script to call UTM repeatedly for the simulation of multicolour data (such as generated in \cite[{Moutou} {et~al.} 2007]{moutou+07}) is available from the author. 
A separate fitting program, UFIT (Universal Fitter) is part of the UTM distribution and may be used to derive best fits to observed data for
any continuously variable parameter.

UTM is
written in the Interactive Data Language (IDL\footnote{IDL is a
product of ITT Visual Information Solutions, http:\/\/www.rsinc.com/idl/}) using object-oriented programming techniques, which requires IDL version 5.1 or higher. The source of UTM and associated
programs is freely distributed under the GNU Public License\footnote{The GNU Public License can be found at http:\/\/www.gnu.org/licenses/gpl.txt} and is available through the author's homepage\footnote{http://www.iac.es/galeria/hdeeg/hdeeghome.html; there follow link to UTM}.

\section{How UTM works}
\begin{figure}[t]
\begin{center}
 \includegraphics[width=10cm]{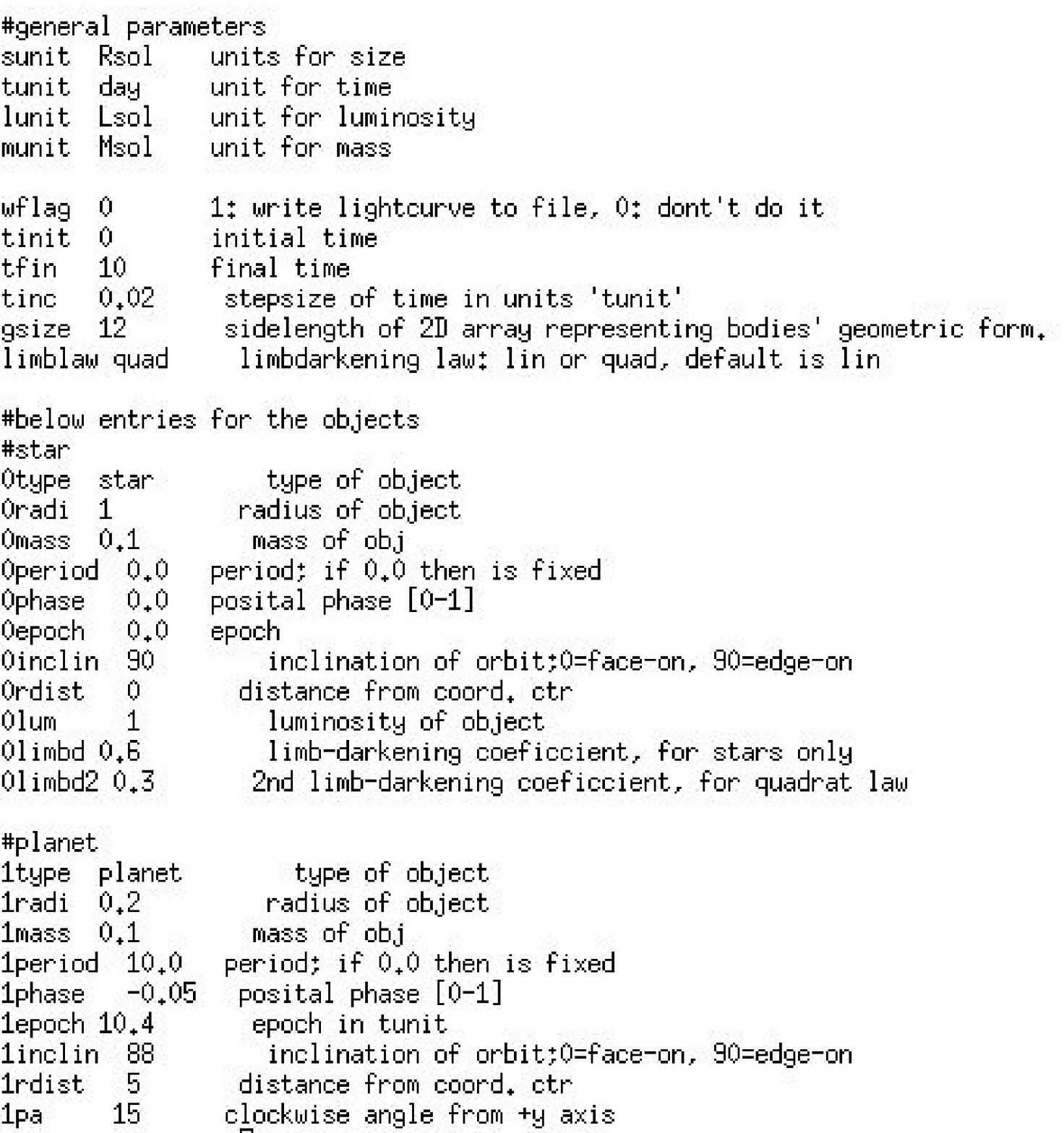} 
 \caption{A simple UTM setup file for a transiting star-planet system. Text behind \# symbols  is ignored, as well as any text behind the second token in a line.}
   \label{fig_setup}
\end{center}
\end{figure}

All input to UTM is given in a setup file in which general parameters and the various bodies with their physical parameters and initial orbital positions are defined in a very simple syntax. A minimalist setup-file for the simulation of a transit of a large planet across a star is shown in Fig.~\ref{fig_setup}. Several further example files may be found in the software distribution of UTM. 
There are two general kinds of bodies: dark
ones (planets, moons, rings) and luminous ones (stars).  For each defined body, a pixelized 2-dimensional representation  in a plane
vertical to the observer (the x-y plane) is created. For all of them, there is a representation of their opacity. For luminous objects (stars), a representation of their surface brightness is
created as well, with the linear, square and quadratic limb-darkening laws
(as defined in \cite[{{Claret} 1998}]{claret98}) being available. While running UTM, an orbital simulation is generated and the 2D representations are being  placed at their corresponding positions into a 3D coordinate system, with the observer being at x,y,z = (0,0,$\infty$). The summed brightness of all luminous objects is then
calculated, taking into account any occultation that may
occur. To do this, UTM checks for each bright object if there is any overlap with another object positioned at  larger z values. If there is one, the overlapping parts of the luminous object's brightness array and of the  occulter's opacity array are regridded to the same scale, with the  coarser (sub)array  being the one that is regridded. This procedure ensures high precision in the resultant brightness even for transits of objects with large size differences, such as Earth-Sun transits. The precision of UTM may also be controlled by defining the basic size of the pixelized representations of the various   objects. Both the precision and the level of visualisations have a strong impact on the computation speed, and a quite mode without any terminal output is also available.

\section{The UFIT fitter}
UFIT is a universal fitting program which can in principle fit any continuously variable parameter to any arbitrary univariate function. This function may be an analytic one or may be generated by some separate algorithm. In the UTM software, UFIT has been implemented to find best fits for \emph{any} continuously variable parameter in a simulation made by UTM. To this end, UFIT reads a UTM-like setup-file that describes an initial model, in which additionally the input light-curve (to which models are being fitted) and the parameters to be fitted have to be specified. UFIT then employs the  Levenberg-Marquardt algorithm, which calls UTM repeatedly with setup-files with slightly varying parameters. From the resultant sets of model light-curves, UFIT derives then partial derivatives of the goodness-of-fit ($\chi^2$ value) against the parameters to be fitted; leading to the best fit in an iterative process. 

\section{Development of UTM}
Since the initial version of UTM, written in early 1999, a number of improvements and additions have been made. These were always compatible with setup-files written for earlier versions, so that results obtained with previous versions can be reproduced reliably in any later one. In the 'traditional' use of UTM, brightness, time and size parameters (object radii, orbital halfaxis) can be given in any units; but this was limited to circular orbits. A recent addition has been a Keplerian code for the generation of either circular or elliptic orbits; this requires however the use of a specific set of units for masses, time and distances (old setup-files, however, still work the same). Current plans for further updates include the addition of star spots as a further class of objects and the possibility to save animations as 'movie-files'. Also foreseen is the inclusion of a further fitting kernel in the UFIT program and the possibility to define constraints among parameters that are being fitted. For the simulation of some more complex objects (e.g. contact binaries), their full 3D representation would be desirable, with a conversion to 2D 'on the fly',  pending on their orientation relative to the observer. Since UTM is released to the public domain, anyone is free to supply further additions to this code; however, the author welcomes suggestions as well. 

\acknowledgement
The author acknowledges a visitorship at the Observatoire de Paris (Meudon) during which the first version of UTM was written, following a suggestion by Jean Schneider.

\end{document}